\documentclass[prl,twocolumn,nofootinbib]{revtex4}

\usepackage{float}
\usepackage{graphicx}
\usepackage{array}
\usepackage{delarray}
\usepackage{amsmath}
\usepackage{amssymb}

\begin{document}

\newcommand{\myvec}[1]{\accentset{\rightharpoonup}{#1}}
\newcommand{\ket}[1]{| #1 \rangle}
\newcommand{\bra}[1]{\langle #1 |}
\newcommand{\op}[1]{{\mathbf #1}}
\newcommand{\ops}[1]{{\boldsymbol #1}}
\newcommand{\Mit}{\mathrm}
\newcommand{\Tr}[2][]{\mathrm{Tr}_{#1} \! \left[ #2 \right]}

\newcommand{\be}{${}^9\mbox{Be}^+$ \:}
\newcommand{\Sstate}{{}^2{\rm S}_{1/2} \:}
\newcommand{\Pone}{{}^2{\rm P}_{1/2} \:}
\newcommand{\Pthree}{{}^2{\rm P}_{3/2} \:}
\newcommand{\Plev}{\mathrm{P}}
\newcommand{\Fd}{F=2, m_F=-2}
\newcommand{\Fu}{F=1, m_F=-1}
\newcommand{\Fs}{{{\mathcal F}_\mathrm{S}}}
\newcommand{\Fone}{{{\mathcal F}_{1/2}}}
\newcommand{\Fthree}{{{\mathcal F}_{3/2}}}
\newcommand{\Fp}{{{\mathcal F}_\mathrm{P}}}
\newcommand{\Fk}{{{\mathcal F}_k}}
\newcommand{\Eps}{{\hat{\epsilon}}}
\newcommand{\phif}{\phi_\Mit{fluct}}

\newcommand{\ua}{{\uparrow}}
\newcommand{\da}{{\downarrow}}
\newcommand{\uan}{{\uparrow^{\mathrm{N}}}}
\newcommand{\dan}{{\downarrow^{\mathrm{N}}}}
\newcommand{\hc}{\mbox{h.c.}}
\newcommand{\cc}{\mbox{c.c.}}
\newcommand{\deltak}{\myvec{\Delta k}}
\newcommand{\omegaud}{\omega_{\da\ua}}
\newcommand{\omegafs}{\omega_{\rm FS}}
\newcommand{\sumj}{\sum_{j=1,2}}
\newcommand{\sumk}{\sum_{k=\{ 1/2,3/2 \}}}
\newcommand{\Omegaud}{\Omega_{\da\ua}}
\newcommand{\deltaud}{\delta_{\da\ua}}
\newcommand{\Oop}{\boldsymbol{\mathcal{O}}}
\newcommand{\nbar}{\overline{n}_\Mit{COM}}
\newcommand{\heat}{\Gamma_\Mit{heat}}

\newcommand{\mydash}{~$\leftrightarrow$~}

\newcommand{\antih}{$\overline{\mbox{H}}$~}
\newcommand{\htwo}{$\mbox{H}_2$ }
\newcommand{\yb}{${}^{174}\mbox{Yb}^+$}

\newcommand{\widtha}{10cm}
\newcommand{\pgwidth}{14cm}
\setlength{\abovedisplayskip}{1mm} 

\title{Laser cooling and trapping with ultrafast pulses}

\author{D. Kielpinski}

\affiliation{Research Laboratory of Electronics and Center for Ultracold Atoms, Massachusetts Institute of Technology,
Cambridge MA 02139}

\begin{abstract}

We propose a new laser cooling method for atomic species whose level structure makes traditional laser cooling difficult. For instance, laser cooling of hydrogen requires vacuum-ultraviolet laser light, while multielectron atoms need laser light at many widely separated frequencies. These restrictions can be eased by laser cooling on two-photon transitions with ultrafast pulse trains. Laser cooling of hydrogen, antihydrogen, and carbon appears feasible, and extension of the technique to molecules may be possible.

\end{abstract}

\maketitle

Laser cooling and trapping are central to modern atomic physics. The low temperatures and long trapping times now routinely achieved by these means have led to great advances in precision spectroscopy and cold collision studies, and provide a suitable starting point for evaporative cooling to Bose-Einstein condensation. However, laser cooling is restricted to less than 20 atomic species, mostly the alkali and alkali-earth metals and the metastable states of noble gases \cite{halbook}.\\ \\

Two obstacles impede the further extension of laser cooling techniques. First, the lowest energy transitions of many atoms, notably hydrogen, lie in the deep UV. Not enough laser power is available in this spectral region to drive effective laser cooling. Second, the complex level structure of many atoms (and all molecules) permits decay of an excited electron into a number of metastable levels widely separated in energy. Each metastable decay channel must typically be repumped by a separate laser, so the laser system becomes unwieldy.\\ \\

These obstacles prevent laser cooling of several interesting atomic species. For instance, efficient laser cooling of hydrogen (H), deuterium (D), and antihydrogen (\antih) would offer impressive gains in atomic spectroscopy, but has remained elusive owing to the lack of power available at the required UV wavelengths. Improved spectroscopy of the 1S -- 2S two-photon transition at 243 nm is the most obvious payoff. This transition plays a unique role in metrology. Measurements of its frequency in H are accurate at the $10^{-14}$ level \cite{hcomb} and assist in determining the value of the Rydberg constant \cite{hconsts}. The isotope shift of the 1S -- 2S transition between H and D gives the most accurately determined value of the D structure radius, tightly constraining nuclear structure calculations \cite{deutshift}. Possibly the most exciting application is a comparison between H and \antih 1S -- 2S frequencies, using the low-energy \antih recently produced at CERN \cite{athena,atrap}. Such comparisons can test CPT symmetry to unprecedented accuracy, probing physics beyond the Standard Model \cite{hbarcpt1,hbarcpt2}. The H 1S -- 2S measurement is currently limited by the $\sim 6$ K temperature of the H beam and can be improved by two orders of magnitude with colder atoms \cite{hcomb}, e.g. in an atomic fountain \cite{hfount}. The \antih formation temperature in the CERN experiments is likely to be of the same order, limiting the corresponding \antih measurement.\\ \\

Cooling of H below a few K currently requires direct contact with superfluid helium \cite{dancryo1,walcryo1}. This method clearly fails for \antih. Attempts to cool D in this way have not been successful \cite{juliasez}, probably because the adsorption energy and recombination rate for D at the liquid helium surface are much higher than for H \cite{donhelium}. Even for H it is cumbersome, requiring a dilution refrigerator and a superconducting magnetic trap, which severely restrict optical access. Current proposals for laser cooling H, D, and \antih involve generation of Lyman $\alpha$ (121 nm) light for excitation of the 1S -- 2P transition. The small amount of light available means that cooling is extremely slow, on the timescale of minutes in the only experiment reported so far \cite{hoptcool}.\\ \\

Atoms with several valence electrons, like carbon, are difficult to laser-cool because of the many widely separated frequencies required for repumping atomic dark states. Spectroscopy on ultracold carbon vapor would greatly improve understanding of chemical bonding between carbon atoms at long range, similar to studies already performed for most alkalis (see \cite{stwparev} for a recent review). Since carbon has such a rich chemistry, this kind of information can potentially impact many fields, from biology to astrophysics. Extending techniques for photoassociative assembly of alkali dimers \cite{stwmolrev} might lead to the controlled generation of small carbon clusters, which exhibit complex structure even for a few atoms \cite{cclusterrev1, cclusterrev2}. The ultracold environment of the clusters may support metastable atomic configurations that are not observable even in supersonic expansions.\\ \\

We propose a scheme for laser cooling and trapping that attacks both obstacles, using pulse trains from ultrafast lasers. The high peak powers of ultrafast pulses enable efficient nonlinear optics far into the UV, greatly increasing the time-averaged optical power available at short wavelengths. At the same time, the many frequencies generated in short pulses can perform the function of repumping lasers, reducing the complexity of laser systems for cooling atoms with multiple valence electrons and possibly allowing laser cooling of molecules. Because of their high peak powers, ultrafast pulses are especially effective for driving two-photon transitions.\\ \\

Laser cooling requires velocity-selective scattering to compress the atomic velocity distribution. A mode-locked pulse train can have high spectral resolution, sufficient to resolve atomic transitions at their natural linewidth \cite{tedspec1,tedspec2,snadden}. The spectrum of such a pulse train is a comb of sharp lines with frequencies $\nu_k = \nu_\Mit{car} + k \nu_\Mit{rep}$, where $k$ is an integer, $\nu_\Mit{car}$ is the optical carrier frequency, and $\nu_\Mit{rep}$ is the pulse repetition rate. If the repetition rate is larger than the atomic linewidth, the atom interacts with only one comb line at a time, so the spectral resolution is just that of a continuous-wave (CW) laser of similar frequency stability. Multiphoton transitions can be resolved in the same way. A broadband source cannot achieve the velocity-selective scattering needed for cooling, so ``white-light" cooling schemes require an additional CW source near each atomic resonance \cite{hoffwhite,whiteexpt1,whiteexpt2}, a difficult requirement in the cases we will consider.\\ \\

In most laser-cooling schemes, each scattering event changes the atomic momentum by one recoil momentum, so the utility of laser cooling depends critically on the scattering rate. We now estimate the efficiencies of CW and pulsed cooling on multiphoton transitions. Consider a $k$-photon transition involving only one intermediate atomic state with linewidth $\Gamma_i$ and saturation intensity $I_\Mit{sat}$, detuned by $\Delta$ from the laser frequency with $\Delta \gg \Gamma_i$. The $k$-photon Rabi frequency $\Omega_k^\Mit{CW}$ for a CW laser of intensity $I$ is then given by \cite{claude}

\begin{equation}
\Omega_k^\Mit{CW} = 2 \eta_\Mit{RWA} \Gamma_i \left( \frac{I}{2 I_\Mit{sat}} \right)^{k/2} \left( \frac{\Gamma_i}{2\pi \Delta}
\right)^{k-1} \label{cwrabi}
\end{equation}

\noindent where $\eta_\Mit{RWA}$ is a constant that accounts for deviations from the rotating-wave approximation at the laser frequency. We can most easily estimate the scattering rate for pulsed excitation in the frequency domain. The spectrum of a train of transform-limited $\mbox{sinc}^2$ pulses with duty cycle $\eta_d \ll 1$ is a series of $N \approx 1/\eta_d$ sharp lines with equal intensities and zero relative phase. For a $k$-photon transition, there are $\approx N^{k-1}$ pathways from the ground to excited state, which all add up coherently. On the other hand, the intensity of each comb line is just $\overline{I}/N$ for a time-averaged light intensity of $\overline{I}$, so the Rabi frequencies in the two cases are related as

\begin{equation}
\Omega_k^\Mit{pulse}(\overline{I}) = N^{k/2-1} \: \Omega_k^\Mit{CW}(I) \label{pulserabi}
\end{equation}
\\ \\

In general, a given laser will produce approximately the same time-averaged power whether it is operated CW or pulsed, so Eq.~(\ref{pulserabi}) predicts that either CW or pulsed operation will drive a two-photon transition at the same rate. Pulsed excitation can be much more efficient for $k>2$, since typically $N \gtrsim 10^3$ for solid-state mode-locked lasers. Since the light shift is a second-order process, a pulsed optical dipole trap has the same properties as a CW dipole trap of the same average intensity. However, the resonant scattering rate for a single-photon transition under pulsed excitation is a factor $N$ smaller than the CW scattering rate, since only one comb line drives the transition.\\ \\

The high spectral resolution of ultrafast pulse trains and their relatively efficient excitation of two-photon transitions suggest that one can use pulse trains to perform laser cooling on two-photon transitions if single-photon cooling is not feasible. In one scenario, the lowest dipole-allowed transition of the atom to be cooled lies at a wavelength $\lesssim 170$ nm, but there is a two-photon transition at $\gtrsim 170$ nm. Continuous-wave light with MHz bandwidth at $\lesssim 170$ nm can usually be generated only by four-wave mixing in atomic vapor \cite{lyagen}. This method is highly technically challenging and yields only tens of nW of radiation. On the other hand, frequency doubling of mode-locked Ti:sapphire laser light can reach near-unit efficiency from infrared to visible \cite{visdbl} and from visible to UV \cite{persaud}, so average powers on the order of 1 W should be achievable for wavelengths $\gtrsim 170$ nm. The huge increase in excitation power can make up for the relative difficulty of exciting a two-photon transition.\\ \\

Pulsed laser cooling on the 1S -- 2S two-photon transition at 243 nm is a good prospect for cooling H, D, and \antih to Doppler-limited temperatures of a few mK. The relevant energy levels are shown in Fig. 1. Although the 2S state is metastable, one can avoid saturating the transition by quenching the 2S state with an applied electric field \cite{zehnle}. The effective linewidth $\Gamma_{\mbox{2S}}$ can be as large as 50 MHz. For weak excitation with CW 243 nm light under maximum quenching, the scattering rate at resonance is $R_2 = 2.8 \times 10^{-7} \: I^2 \: \mbox{Hz}\: \mbox{W}^{-2}\: \mbox{cm}^4$ \cite{sandberg}. Since $\eta_\Mit{RWA}$ in Eq. \ref{cwrabi} varies slowly with frequency, the equivalence $\Omega_2^\Mit{CW} = \Omega_2^\Mit{pulse}$ remains valid as long as the pulse bandwidth is much less than the optical carrier frequency.\\ \\

\begin{figure}
\begin{center}
\includegraphics*[width=6cm]{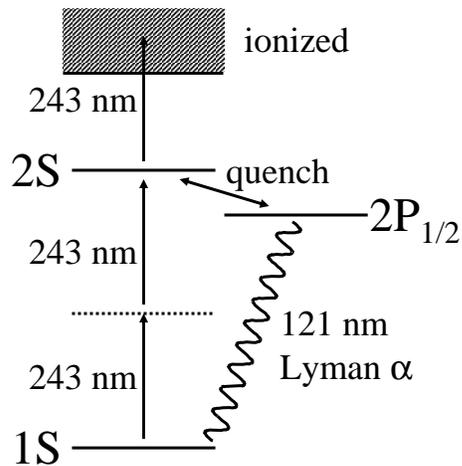}
\caption{Energy level diagram for hydrogen cooling. The 243 nm light excites the atoms to the 2S state, which is quenched by a static electric field to the 2P state. The atoms reradiate on the 1S--2P transition at 121 nm. While in the 2S state, an atom can be photoionized by a single 243 nm photon.}
\end{center}
\end{figure}

The upper limit to the usable intensity comes from one-photon ionization of the 2S state. The photoionization rate, $R_\Mit{PI} = 11.4 \: I \: \mbox{Hz}\: \mbox{W}^{-1} \: \mbox{cm}^2$ \cite{sandberg}, is the same for pulsed and CW excitation; since the final state is a continuum with slowly varying matrix element, all comb lines contribute equally. If an atom undergoes $N_\gamma$ scattering events in cooling, we require $R_\Mit{PI}/\Gamma_\Mit{2S} \ll N_\gamma$ to avoid photoionization, so the maximum quenching is indeed desirable. A single scattering event changes the velocity of an H atom by $3.25 \:\mbox{m}/\mbox{s}$, so about half the H atoms in a 6 K gas are cooled to the Doppler limit after 100 scattering events. If 90\% of the atoms are allowed to photoionize, the desired intensity is $\sim 100 \:\mbox{kW}/\mbox{cm}^2$ and the scattering rate is $\sim 2.8$ kHz.\\ \\

Intensities of this order require the high UV powers available from ultrafast pulse trains. A possible scenario might involve doubling a mode-locked Ti:S pulse train twice. The near unit doubling efficiencies from IR to UV \cite{visdbl,persaud} enable generation of $\sim 1$ W of power at 243 nm. Coupling this light into a resonant enhancement cavity with a buildup factor of $\sim 30$ \cite{ferdi1,dana1} could achieve a scattering rate of a few kHz over a beam diameter $\sim 300 \:\mu$m. While this scheme is clearly less efficient than laser cooling of alkali atoms, it is competitive with other methods for laser-cooling H. Currently only $\sim 20$ mW of CW 243 nm light is usually generated \cite{sandberg,ferdi1}, so the scattering rate drops to $\sim 1$ Hz for CW two-photon cooling. Pulsed two-photon cooling compares well to cooling on the 121 nm 1S -- 2P transition owing to the technical difficulties of generating 121 nm light. The first 121 nm sources were developed over 20 years ago \cite{firstlya}, but the highest CW power reported is still only 20 nW \cite{lyagen}. The scattering rate for pulsed two-photon cooling is approximately equal to that for 121 nm cooling, given $\sim 300 \:\mu$m beam diameters in both cases.\\ \\

The use of pulsed lasers simplifies laser cooling of atoms with multiple valence electrons. Because these atoms typically have many low-lying metastable states, they are optically pumped into a dark state by the cooling light unless all metastable states are excited simultaneously. Efficient cooling requires velocity-selective excitation of all transitions, so a narrowband radiation source must address each transition. This task is difficult for CW lasers, but a single pulsed laser is sufficient. The octave-spanning laser oscillators currently available \cite{bartoctave} can easily cover the entire spectral range needed for excitation of all transitions. Although the transitions are spaced more or less randomly with respect to the comb of frequencies generated by the pulse train, the gaps between transition and laser frequencies are smaller than the repetition rate and can easily be spanned by an electro-optic modulator driven at MHz to GHz frequencies.\\ \\

Such an RF-modulated pulse train might be used for laser cooling of carbon. In carbon, the wavelengths of the lowest dipole-allowed transitions lie blue of 170 nm, so one-photon cooling is no easier than for hydrogen. There are six states in the ground $2s^2 \: 2p^2$ electronic configuration, all having radiative lifetimes $> 1$ s and spanning the energy range $-90820$ to $-69172 \:\mbox{cm}^{-1}$ (taking the ionization limit as $0 \:\mbox{cm}^{-1}$). One-photon cooling thus would require six vacuum UV lasers, a formidable technical challenge. However, carbon has many two-photon transitions out of the ground-state manifold that can be excited with light in the 200 -- 300 nm range \cite{nistdata}. Second-order perturbation theory suggests transition rates of $10^{-3}$ to $10^{-5} \:\: I^2\:\: \mbox{Hz} \:\: \mbox{W}^{-2} \:\: \mbox{cm}^4$, orders of magnitude higher than for hydrogen 1S -- 2S. If high-lying excited states, within $10000 \: \mbox{cm}^{-1}$ of ionization, are chosen, excited-state photoionization can also be orders of magnitude smaller than for hydrogen 1S -- 2S \cite{opacity}. On the other hand, the recoil velocity and radiative lifetime both decrease an order of magnitude. Because six transitions must be driven, the power available to drive each transition decreases a factor of six, while resonant enhancement of cooling power becomes impractical. These advantages and disadvantages roughly balance for realistic parameter values, so laser cooling of carbon from few K to few mK temperatures appears feasible.\\ \\

The cases of hydrogen and carbon suggest that pulsed two-photon excitation can efficiently cool a variety of atomic species to mK temperatures if the atoms are precooled to a few K. A variety of atomic and molecular gases have been cooled to these temperatures by thermalization with helium buffer gas \cite{bufferrev}. To obtain monatomic gases of refractory elements like carbon, one typically uses a hollow cathode discharge beam \cite{cbeam} which operates at high temperature. Buffer-gas cooling of such a beam, along the lines of \cite{bufferbeam}, might provide a quite general precooling method for subsequent pulsed two-photon cooling.\\ \\

Pulsed two-photon excitation might also be useful in the laser cooling of molecules. The tens or hundreds of metastable levels in a typical diatomic molecule must all be excited for scattering to take place. To keep these transitions resolved, the repetition rate of the laser must become very high and the power available for each transition becomes small. Although cooling with a RF-modulated wideband pulse train becomes ineffective in this case, a relatively narrowband picosecond pulse train can generate Raman sidebands from Doppler-free transitions in a molecular vapor, which can address the molecular vibrational levels independently for efficient cooling \cite{msfcool}.\\ \\

We have presented a new method of laser cooling based on two-photon excitation with ultrafast pulse trains. Pulse trains can provide the velocity selection necessary for laser cooling, and pulsed light excites two-photon transitions as efficiently as CW laser light of the same average intensity. Frequency conversion is more efficient for ultrafast pulses, giving them an advantage for two-photon laser cooling of atoms whose lowest transitions lie in the vacuum UV, such as H and \antih. It also seems possible to cool multielectron atoms, for instance carbon, by modulating a single pulse train at radio frequencies. In combination with buffer-gas precooling \cite{bufferrev,bufferbeam}, this method offers the chance to produce mK samples of a variety of new atomic species. The application of similar techniques to laser cooling of molecules remains a question for future work.\\ \\

\begin{acknowledgments}
The author would like to acknowledge the great help afforded by consultation with Prof. D. Kleppner and J. Steinberger about ultracold hydrogen, Prof. F.X. K\"artner about ultrafast lasers, and Prof. J. Doyle and Z. Stone about buffer-gas cooling, as well as discussions on laser cooling with Profs. D.E. Pritchard and W. Ketterle and insightful comments on the manuscript by Dr. D. Schneble. The author was financially supported by a Pappalardo Fellowship.
\end{acknowledgments}

\bibliography{kielpinski_bib}

\begin{thebibliography}{40}
\expandafter\ifx\csname natexlab\endcsname\relax\def\natexlab#1{#1}\fi
\expandafter\ifx\csname bibnamefont\endcsname\relax
  \def\bibnamefont#1{#1}\fi
\expandafter\ifx\csname bibfnamefont\endcsname\relax
  \def\bibfnamefont#1{#1}\fi
\expandafter\ifx\csname citenamefont\endcsname\relax
  \def\citenamefont#1{#1}\fi
\expandafter\ifx\csname url\endcsname\relax
  \def\url#1{\texttt{#1}}\fi
\expandafter\ifx\csname urlprefix\endcsname\relax\def\urlprefix{URL }\fi
\providecommand{\bibinfo}[2]{#2}
\providecommand{\eprint}[2][]{\url{#2}}

\bibitem[{\citenamefont{Metcalf and van~der Straten}(1999)}]{halbook}
\bibinfo{author}{\bibfnamefont{H.}~\bibnamefont{Metcalf}} \bibnamefont{and}
  \bibinfo{author}{\bibfnamefont{P.}~\bibnamefont{van~der Straten}},
  \emph{\bibinfo{title}{Laser Cooling and Trapping}}
  (\bibinfo{publisher}{Springer}, \bibinfo{address}{New York},
  \bibinfo{year}{1999}).

\bibitem[{\citenamefont{Niering et~al.}(2000)}]{hcomb}
\bibinfo{author}{\bibfnamefont{M.}~\bibnamefont{Niering}} \bibnamefont{et~al.},
  \bibinfo{journal}{Phys. Rev. Lett.} \textbf{\bibinfo{volume}{84}},
  \bibinfo{pages}{5496} (\bibinfo{year}{2000}).

\bibitem[{\citenamefont{Schwob et~al.}(1999)}]{hconsts}
\bibinfo{author}{\bibfnamefont{C.}~\bibnamefont{Schwob}} \bibnamefont{et~al.},
  \bibinfo{journal}{Phys. Rev. Lett.} \textbf{\bibinfo{volume}{82}},
  \bibinfo{pages}{4960} (\bibinfo{year}{1999}).

\bibitem[{\citenamefont{Huber et~al.}(1998)}]{deutshift}
\bibinfo{author}{\bibfnamefont{A.}~\bibnamefont{Huber}} \bibnamefont{et~al.},
  \bibinfo{journal}{Phys. Rev. Lett.} \textbf{\bibinfo{volume}{80}},
  \bibinfo{pages}{468} (\bibinfo{year}{1998}).

\bibitem[{\citenamefont{Amoretti et~al.}(2002)}]{athena}
\bibinfo{author}{\bibfnamefont{M.}~\bibnamefont{Amoretti}}
  \bibnamefont{et~al.}, \bibinfo{journal}{Nature}
  \textbf{\bibinfo{volume}{419}}, \bibinfo{pages}{456} (\bibinfo{year}{2002}).

\bibitem[{\citenamefont{Gabrielse et~al.}(in press)}]{atrap}
\bibinfo{author}{\bibfnamefont{G.}~\bibnamefont{Gabrielse}}
  \bibnamefont{et~al.}, \bibinfo{journal}{Phys. Rev. Lett.}  (\bibinfo{year}{in
  press}).

\bibitem[{\citenamefont{Bluhm et~al.}(1999)\citenamefont{Bluhm, Kosteleck\'y,
  and Russell}}]{hbarcpt1}
\bibinfo{author}{\bibfnamefont{R.}~\bibnamefont{Bluhm}},
  \bibinfo{author}{\bibfnamefont{V.}~\bibnamefont{Kosteleck\'y}},
  \bibnamefont{and} \bibinfo{author}{\bibfnamefont{N.}~\bibnamefont{Russell}},
  \bibinfo{journal}{Phys. Rev. Lett.} \textbf{\bibinfo{volume}{82}},
  \bibinfo{pages}{2254} (\bibinfo{year}{1999}).

\bibitem[{\citenamefont{Gabrielse}(2000)}]{hbarcpt2}
\bibinfo{author}{\bibfnamefont{G.}~\bibnamefont{Gabrielse}},
  \bibinfo{journal}{Adv. At. Mol. Opt. Phys.} \textbf{\bibinfo{volume}{45}},
  \bibinfo{pages}{1} (\bibinfo{year}{2000}).

\bibitem[{\citenamefont{Beausoleil and H\"ansch}(1986)}]{hfount}
\bibinfo{author}{\bibfnamefont{R.}~\bibnamefont{Beausoleil}} \bibnamefont{and}
  \bibinfo{author}{\bibfnamefont{T.}~\bibnamefont{H\"ansch}},
  \bibinfo{journal}{Phys. Rev. A} \textbf{\bibinfo{volume}{33}},
  \bibinfo{pages}{1661} (\bibinfo{year}{1986}).

\bibitem[{\citenamefont{Hess et~al.}(1987)}]{dancryo1}
\bibinfo{author}{\bibfnamefont{H.}~\bibnamefont{Hess}} \bibnamefont{et~al.},
  \bibinfo{journal}{Phys. Rev. Lett.} \textbf{\bibinfo{volume}{59}},
  \bibinfo{pages}{672} (\bibinfo{year}{1987}).

\bibitem[{\citenamefont{van Roijen et~al.}(1988)}]{walcryo1}
\bibinfo{author}{\bibfnamefont{R.}~\bibnamefont{van Roijen}}
  \bibnamefont{et~al.}, \bibinfo{journal}{Phys. Rev. Lett.}
  \textbf{\bibinfo{volume}{61}}, \bibinfo{pages}{931} (\bibinfo{year}{1988}).

\bibitem[{\citenamefont{Steinberger}(2003)}]{juliasez}
\bibinfo{author}{\bibfnamefont{J.}~\bibnamefont{Steinberger}},
  \bibinfo{type}{private communication} (\bibinfo{year}{2003}).

\bibitem[{\citenamefont{Mosk et~al.}(2001)\citenamefont{Mosk, Reynolds, and
  Hijmans}}]{donhelium}
\bibinfo{author}{\bibfnamefont{A.}~\bibnamefont{Mosk}},
  \bibinfo{author}{\bibfnamefont{M.}~\bibnamefont{Reynolds}}, \bibnamefont{and}
  \bibinfo{author}{\bibfnamefont{T.}~\bibnamefont{Hijmans}},
  \bibinfo{journal}{Phys. Rev. A} \textbf{\bibinfo{volume}{64}},
  \bibinfo{pages}{022901} (\bibinfo{year}{2001}).

\bibitem[{\citenamefont{Setija et~al.}(1993)}]{hoptcool}
\bibinfo{author}{\bibfnamefont{I.}~\bibnamefont{Setija}} \bibnamefont{et~al.},
  \bibinfo{journal}{Phys. Rev. Lett.} \textbf{\bibinfo{volume}{70}},
  \bibinfo{pages}{2257} (\bibinfo{year}{1993}).

\bibitem[{\citenamefont{Stwalley and Wang}(1999)}]{stwparev}
\bibinfo{author}{\bibfnamefont{W.}~\bibnamefont{Stwalley}} \bibnamefont{and}
  \bibinfo{author}{\bibfnamefont{H.}~\bibnamefont{Wang}}, \bibinfo{journal}{J.
  Mol. Spec.} \textbf{\bibinfo{volume}{195}}, \bibinfo{pages}{194}
  (\bibinfo{year}{1999}).

\bibitem[{\citenamefont{Bahns et~al.}(2000)\citenamefont{Bahns, Gould, and
  Stwalley}}]{stwmolrev}
\bibinfo{author}{\bibfnamefont{J.}~\bibnamefont{Bahns}},
  \bibinfo{author}{\bibfnamefont{P.}~\bibnamefont{Gould}}, \bibnamefont{and}
  \bibinfo{author}{\bibfnamefont{W.}~\bibnamefont{Stwalley}}
  (\bibinfo{publisher}{Academic Press}, \bibinfo{year}{2000}),
  vol.~\bibinfo{volume}{42} of \emph{\bibinfo{series}{Adv. At. Mol. Opt.
  Phys.}}, p. \bibinfo{pages}{171}.

\bibitem[{\citenamefont{W.~Weltner and Zee}(1989)}]{cclusterrev1}
\bibinfo{author}{\bibfnamefont{J.}~\bibnamefont{W.~Weltner}} \bibnamefont{and}
  \bibinfo{author}{\bibfnamefont{R.~V.} \bibnamefont{Zee}},
  \bibinfo{journal}{Chem. Rev.} \textbf{\bibinfo{volume}{89}},
  \bibinfo{pages}{1713} (\bibinfo{year}{1989}).

\bibitem[{\citenamefont{Orden and Saykally}(1998)}]{cclusterrev2}
\bibinfo{author}{\bibfnamefont{A.~V.} \bibnamefont{Orden}} \bibnamefont{and}
  \bibinfo{author}{\bibfnamefont{R.}~\bibnamefont{Saykally}},
  \bibinfo{journal}{Chem. Rev.} \textbf{\bibinfo{volume}{98}},
  \bibinfo{pages}{2313} (\bibinfo{year}{1998}).

\bibitem[{\citenamefont{Teets et~al.}(1977)\citenamefont{Teets, Eckstein, and
  H\"ansch}}]{tedspec1}
\bibinfo{author}{\bibfnamefont{R.}~\bibnamefont{Teets}},
  \bibinfo{author}{\bibfnamefont{J.}~\bibnamefont{Eckstein}}, \bibnamefont{and}
  \bibinfo{author}{\bibfnamefont{T.~W.} \bibnamefont{H\"ansch}},
  \bibinfo{journal}{Phys. Rev. Lett.} \textbf{\bibinfo{volume}{38}},
  \bibinfo{pages}{760} (\bibinfo{year}{1977}).

\bibitem[{\citenamefont{Eckstein et~al.}(1978)\citenamefont{Eckstein, Ferguson,
  and H\"ansch}}]{tedspec2}
\bibinfo{author}{\bibfnamefont{J.}~\bibnamefont{Eckstein}},
  \bibinfo{author}{\bibfnamefont{A.}~\bibnamefont{Ferguson}}, \bibnamefont{and}
  \bibinfo{author}{\bibfnamefont{T.~W.} \bibnamefont{H\"ansch}},
  \bibinfo{journal}{Phys. Rev. Lett.} \textbf{\bibinfo{volume}{40}},
  \bibinfo{pages}{847} (\bibinfo{year}{1978}).

\bibitem[{\citenamefont{Snadden et~al.}(1996)}]{snadden}
\bibinfo{author}{\bibfnamefont{M.}~\bibnamefont{Snadden}} \bibnamefont{et~al.},
  \bibinfo{journal}{Opt. Comm.} \textbf{\bibinfo{volume}{125}},
  \bibinfo{pages}{70} (\bibinfo{year}{1996}).

\bibitem[{\citenamefont{Hoffnagle}(1988)}]{hoffwhite}
\bibinfo{author}{\bibfnamefont{J.}~\bibnamefont{Hoffnagle}},
  \bibinfo{journal}{Opt. Lett.} \textbf{\bibinfo{volume}{13}},
  \bibinfo{pages}{102} (\bibinfo{year}{1988}).

\bibitem[{\citenamefont{Littler et~al.}(1991)}]{whiteexpt1}
\bibinfo{author}{\bibfnamefont{I.}~\bibnamefont{Littler}} \bibnamefont{et~al.},
  \bibinfo{journal}{Z. Phys. D} \textbf{\bibinfo{volume}{18}},
  \bibinfo{pages}{307} (\bibinfo{year}{1991}).

\bibitem[{\citenamefont{Zhu et~al.}(1991)\citenamefont{Zhu, Oates, and
  Hall}}]{whiteexpt2}
\bibinfo{author}{\bibfnamefont{M.}~\bibnamefont{Zhu}},
  \bibinfo{author}{\bibfnamefont{C.}~\bibnamefont{Oates}}, \bibnamefont{and}
  \bibinfo{author}{\bibfnamefont{J.}~\bibnamefont{Hall}},
  \bibinfo{journal}{Phys. Rev. Lett.} \textbf{\bibinfo{volume}{67}},
  \bibinfo{pages}{46} (\bibinfo{year}{1991}).

\bibitem[{\citenamefont{Cohen-Tannoudji
  et~al.}(1992)\citenamefont{Cohen-Tannoudji, Dupont-Roc, and
  Grynberg}}]{claude}
\bibinfo{author}{\bibfnamefont{C.}~\bibnamefont{Cohen-Tannoudji}},
  \bibinfo{author}{\bibfnamefont{J.}~\bibnamefont{Dupont-Roc}},
  \bibnamefont{and} \bibinfo{author}{\bibfnamefont{G.}~\bibnamefont{Grynberg}},
  \emph{\bibinfo{title}{Atom-Photon Interactions}}
  (\bibinfo{publisher}{Wiley-Interscience}, \bibinfo{address}{New York},
  \bibinfo{year}{1992}).

\bibitem[{\citenamefont{Eikema et~al.}(1999)\citenamefont{Eikema, Walz, and
  H\"ansch}}]{lyagen}
\bibinfo{author}{\bibfnamefont{K.}~\bibnamefont{Eikema}},
  \bibinfo{author}{\bibfnamefont{J.}~\bibnamefont{Walz}}, \bibnamefont{and}
  \bibinfo{author}{\bibfnamefont{T.}~\bibnamefont{H\"ansch}},
  \bibinfo{journal}{Phys. Rev. Lett.} \textbf{\bibinfo{volume}{83}},
  \bibinfo{pages}{3828} (\bibinfo{year}{1999}).

\bibitem[{\citenamefont{Nebel and Beigang}(1991)}]{visdbl}
\bibinfo{author}{\bibfnamefont{A.}~\bibnamefont{Nebel}} \bibnamefont{and}
  \bibinfo{author}{\bibfnamefont{R.}~\bibnamefont{Beigang}},
  \bibinfo{journal}{Opt. Lett.} \textbf{\bibinfo{volume}{16}},
  \bibinfo{pages}{1729} (\bibinfo{year}{1991}).

\bibitem[{\citenamefont{Persaud et~al.}(1978)\citenamefont{Persaud, Tolchard,
  and Ferguson}}]{persaud}
\bibinfo{author}{\bibfnamefont{M.}~\bibnamefont{Persaud}},
  \bibinfo{author}{\bibfnamefont{J.}~\bibnamefont{Tolchard}}, \bibnamefont{and}
  \bibinfo{author}{\bibfnamefont{A.}~\bibnamefont{Ferguson}},
  \bibinfo{journal}{IEEE J. Quant. Elec.} \textbf{\bibinfo{volume}{40}},
  \bibinfo{pages}{847} (\bibinfo{year}{1978}).

\bibitem[{\citenamefont{Zehnl\'{e} and Garreau}(2001)}]{zehnle}
\bibinfo{author}{\bibfnamefont{V.}~\bibnamefont{Zehnl\'{e}}} \bibnamefont{and}
  \bibinfo{author}{\bibfnamefont{J.}~\bibnamefont{Garreau}},
  \bibinfo{journal}{Phys. Rev. A} \textbf{\bibinfo{volume}{63}},
  \bibinfo{pages}{021402} (\bibinfo{year}{2001}).

\bibitem[{\citenamefont{Sandberg}(1993)}]{sandberg}
\bibinfo{author}{\bibfnamefont{J.}~\bibnamefont{Sandberg}}, Ph.D. thesis,
  \bibinfo{school}{Massachusetts Institute of Technology}
  (\bibinfo{year}{1993}).

\bibitem[{\citenamefont{Schmidt-Kaler et~al.}(1995)}]{ferdi1}
\bibinfo{author}{\bibfnamefont{F.}~\bibnamefont{Schmidt-Kaler}}
  \bibnamefont{et~al.}, \bibinfo{journal}{Phys. Rev. A}
  \textbf{\bibinfo{volume}{51}}, \bibinfo{pages}{2789} (\bibinfo{year}{1995}).

\bibitem[{\citenamefont{Berkeland et~al.}(1995)\citenamefont{Berkeland, Hinds,
  and Boshier}}]{dana1}
\bibinfo{author}{\bibfnamefont{D.}~\bibnamefont{Berkeland}},
  \bibinfo{author}{\bibfnamefont{E.}~\bibnamefont{Hinds}}, \bibnamefont{and}
  \bibinfo{author}{\bibfnamefont{M.}~\bibnamefont{Boshier}},
  \bibinfo{journal}{Phys. Rev. Lett.} \textbf{\bibinfo{volume}{75}},
  \bibinfo{pages}{2470} (\bibinfo{year}{1995}).

\bibitem[{\citenamefont{Mahon et~al.}(1978)\citenamefont{Mahon, McIlrath, and
  Koopman}}]{firstlya}
\bibinfo{author}{\bibfnamefont{R.}~\bibnamefont{Mahon}},
  \bibinfo{author}{\bibfnamefont{T.}~\bibnamefont{McIlrath}}, \bibnamefont{and}
  \bibinfo{author}{\bibfnamefont{D.}~\bibnamefont{Koopman}},
  \bibinfo{journal}{Appl. Phys. Lett.} \textbf{\bibinfo{volume}{33}},
  \bibinfo{pages}{305} (\bibinfo{year}{1978}).

\bibitem[{\citenamefont{Bartels and Kurz}(2002)}]{bartoctave}
\bibinfo{author}{\bibfnamefont{A.}~\bibnamefont{Bartels}} \bibnamefont{and}
  \bibinfo{author}{\bibfnamefont{H.}~\bibnamefont{Kurz}},
  \bibinfo{journal}{Opt. Lett.} \textbf{\bibinfo{volume}{27}},
  \bibinfo{pages}{1839} (\bibinfo{year}{2002}).

\bibitem[{\citenamefont{Martin et~al.}(2002)}]{nistdata}
\bibinfo{author}{\bibfnamefont{W.}~\bibnamefont{Martin}} \bibnamefont{et~al.},
  \bibinfo{type}{NIST Atomic Spectra Database (version 2.0), available at
  http://physics.nist.gov/asd}, \bibinfo{institution}{National Institute of
  Standards and Technology}, \bibinfo{address}{Gaithersburg, MD}
  (\bibinfo{year}{2002}).

\bibitem[{opa(2002)}]{opacity}
\bibinfo{type}{Opacity Project TOPbase Database, available at
  http://heasarc.gsfc.nasa.gov/topbase} (\bibinfo{year}{2002}).

\bibitem[{\citenamefont{deCarvalho et~al.}(1999)}]{bufferrev}
\bibinfo{author}{\bibfnamefont{R.}~\bibnamefont{deCarvalho}}
  \bibnamefont{et~al.}, \bibinfo{journal}{Eur. Phys. J. D}
  \textbf{\bibinfo{volume}{7}}, \bibinfo{pages}{289} (\bibinfo{year}{1999}).

\bibitem[{\citenamefont{O'Brian and Lawler}(1997)}]{cbeam}
\bibinfo{author}{\bibfnamefont{T.~R.} \bibnamefont{O'Brian}} \bibnamefont{and}
  \bibinfo{author}{\bibfnamefont{J.~E.} \bibnamefont{Lawler}},
  \bibinfo{journal}{J. Quant. Spectrosc. Radiat. Transfer}
  \textbf{\bibinfo{volume}{57}}, \bibinfo{pages}{309} (\bibinfo{year}{1997}).

\bibitem[{\citenamefont{Egorov et~al.}(2002)}]{bufferbeam}
\bibinfo{author}{\bibfnamefont{D.}~\bibnamefont{Egorov}} \bibnamefont{et~al.},
  \bibinfo{journal}{Phys. Rev. A} \textbf{\bibinfo{volume}{66}},
  \bibinfo{pages}{043401} (\bibinfo{year}{2002}).

\bibitem[{\citenamefont{Bahns et~al.}(1996)\citenamefont{Bahns, Stwalley, and
  Gould}}]{msfcool}
\bibinfo{author}{\bibfnamefont{J.~T.} \bibnamefont{Bahns}},
  \bibinfo{author}{\bibfnamefont{W.~C.} \bibnamefont{Stwalley}},
  \bibnamefont{and} \bibinfo{author}{\bibfnamefont{P.~L.} \bibnamefont{Gould}},
  \bibinfo{journal}{J. Chem. Phys.} \textbf{\bibinfo{volume}{104}},
  \bibinfo{pages}{9689} (\bibinfo{year}{1996}).

\end{thebibliography}

\end{document}